\documentstyle[aps,prl,multicol,epsfig]{revtex}
\begin{document}
\title{ \bf  Ginsburg-Landau theory of supersolid}
\author{ \bf  Jinwu Ye  }
\address{ Department of Physics, The Pennsylvania State University, University Park, PA, 16802 }
\date{\today}
\maketitle
\begin{abstract}
    We develop a simple Ginsburg-Landau theory to
    study all the possible phases and phase transitions in $^{4}He $,
    analyze the condition for the existence of the supersolid (SS)
    and map out its global phase diagram from a unified framework.
    If the condition favors the existence of the SS,
    we use the GL theory to address several experimental
    facts and also make some predictions that are amenable to
    experimental tests. A key prediction is that the X-ray
    scattering intensity from the SS ought to have an additional modulation
    over that of the NS. The modulation amplitude is proportional
    to the Non-Classical Rotational-Inertial (NCRI) observed in the
    torsional oscillator experiments.
\end{abstract}

\begin{multicols}{2}
{\bf 1. Introduction: }
     A solid can not flow. While a superfluid can flow without any resistance.
     A supersolid (SS) is a new state of matter which has both the
     solid and superfluid order. The possibility of a supersolid phase in $^{4}He $ due to very large zero point
     quantum fluctuations  was
     theoretically speculated in 1970 \cite{and,ches,leg,sas}. Over the last 35
     years, a number of experiments have been designed to search for
     the supersolid state without success. However, recently, by
     using torsional oscillator measurement, a PSU group lead by Chan observed a marked
     $ 1 \sim 2 \% $ Non-Classical Rotational Inertial ( NCRI ) of solid $^{4}He $ at $ \sim 0.2 K $,
     both when embedded in Vycor glass and in bulk $^{4}He $ \cite{chan}.
     The NCRI is a low temperature reduction in the rotational moment of inertia
     due to the superfluid component of solid $^{4}He $ \cite{leg}.
     The authors suggested that the supersolid state is responsible
     for the NCRI.
    The PSU experiments rekindled extensive theoretical interests \cite{micro,ander,dorsey} in the still controversial
    supersolid phase of $^{4}He $. There are two kinds of complementary theoretical approaches. The first is the microscopic
    numerical simulation \cite{micro}. The second is the phenomenological
    approach \cite{ander,dorsey}. In this paper, by constructing a Ginsburg
    Landau ( GL ) theory, we will address the
    following two questions  : (1) What is the condition for the existence
    of the SS state ? (2) If the SS exists, what are the properties of the
    supersolid to be tested by possible new experiments.
    We develop a simple GL theory  to map out the $^{4}He $ phase diagram  and
    study all the phases and phase transitions in a unified
    framework. If the repulsive coupling in Eqn.\ref{int} in the GL theory is sufficiently small,
    the SS phase becomes stable at low enough temperature.
    The resulting solid at high pressure is an incommensurate solid
    with zero point vacancies whose condensation leads to the
    formation of the SS. The theory  can be used to address several phenomena
    observed in the PSU experiments and also make sharp predictions to be tested by possible future
    experiments, especially X-ray scattering experiments in the SS state.

{\bf 2. Ginsburg-Landau theory of $^{4}He $:}  Let's start by
     reviewing all the known phases in $^{4}He $. The density of a normal solid (NS) is
     defined as $ n( \vec{x}
     )=  n_{0}  + \sum^{\prime}_{\vec{G}} n_{\vec{G}} e^{i \vec{G} \cdot
     \vec{x} } $ where $ n^{*}_{\vec{G}}=n_{- \vec{G}}$ and $ \vec{G} $ is any non-zero reciprocal lattice vector.
     In a normal liquid (NL), if the static liquid structure factor $ S(k) $ has its first maximum peak
     at $ \vec{k}_{n} $, then near $ k_{n} $, $ S^{-1}(k) \sim  r_{n} + c ( k^{2}-k^{2}_{n} )^{2}  $.
     If the liquid-solid transition is weakly first order, it is known that
     the classical free energy to describe the NL-NS transition is \cite{tom}:
\begin{eqnarray}
  f_{n} & = &  \sum_{\vec{G}} \frac{1}{2}  r_{\vec{G}} | n_{\vec{G}} |^{2}
    -w  \sum_{\vec{G}_1,\vec{G}_2,\vec{G}_3} n_{\vec{G}_1}
    n_{\vec{G}_2} n_{\vec{G}_3} \delta_{ \vec{G}_1 + \vec{G}_2 +
    \vec{G}_3,0 }  \nonumber   \\
    & + &  u \sum_{\vec{G}_1,\vec{G}_2,\vec{G}_3,\vec{G}_4 } n_{\vec{G}_1}
    n_{\vec{G}_2} n_{\vec{G}_3} n_{ \vec{G}_4 }  \delta_{ \vec{G}_1 + \vec{G}_2 +
    \vec{G}_3+ \vec{G}_4, 0 } + \cdots
\label{sl}
\end{eqnarray}
    where $ r_{\vec{G}}=r_{n} + c ( G^{2}-k^{2}_{n} )^{2} $ is the
    tuning parameter controlled by the pressure.

    It was known that the Superfluid (SF) to Normal Liquid transition  at finite temperature is a 3d
    XY transition described by:
\begin{equation}
  f_{\psi}  =  K | \nabla \psi |^{2} + t | \psi |^{2} +  u  |\psi |^{4}  + \cdots
\label{sfl}
\end{equation}
   where $ \psi $ is the complex order parameter and $ t $ is the
   tuning parameter controlled by the temperature.

    The coupling between $ n ( \vec{x} ) $ and $ \psi( \vec{x} ) $ consistent
    with all the symmetry can be written down as:
\begin{equation}
  f_{int} =   v_{1} n ( \vec{x} ) | \psi( \vec{x} ) |^{2} +   \cdots
\label{int}
\end{equation}
   where the interaction must be repulsive $ v_{1} > 0
   $. Due to the lack of particle-hole symmetry in the NS,
   additional terms like  $ n( \vec{x} ) \psi^{\dagger} \partial_{\tau} \psi $ should exist and
   is very important at zero temperature
   and will be investigated in \cite{un}. However, in the classical
   phase transitions investigated in this paper, this term can still
   be neglected. As shown in section 5, the repulsive interaction is  proportional to
   the temperature shift $ T_{SF}-T_{SS} $ in Fig.1.

     In an effective GL theory, $ n( \vec{x} ) $ and $ \psi (\vec{x} ) $
     emerge as two independent order parameters.  The total density of the system is $ n_{t}(x)= n(x) +
     |\psi(x)|^{2} $ where $ n(x) $ is the normal density and $  |\psi(x)|^{2} $ is the superfluid density.
     A NS is defined by $ n_{ \vec{G} } \neq 0,
     < \psi> =0 $, while a SS is defined by $
      n_{ \vec{G} } \neq 0, ~< \psi> \neq 0 $.
      From the NL side, one can approach both the NS and the SF.
      Inside the NL,  $ t > 0 $, $ \psi $ has a gap, so can be integrated
      out, we recover the NS-NL transition tuned by $ r_{\vec{G}} $ in Eqn.\ref{sl} ( Fig.1 ).
      Inside the NL $ < n( \vec{x} )> =n_{0} $, so we can set
      $ n_{ \vec{G} }=0 $ for $ \vec{G} \neq 0 $ in Eqn.\ref{int},
      then we recover the NL to SF transition tuned by $ t $ in
      Eqn.\ref{sfl} ( Fig.1 ).

     Although the NL-NS and NL-SF transitions are well understood,
     so far, the SF-NS transition has not been investigated seriously. This
     transition may be in a completely different universality class than the
     NL-NS transition, because both sides break two completely
     different symmetry: internal global $ U(1) $ symmetry and
     translational symmetry. It is possible that the solid reached from the SF side is a new
     kind of solid than the NS reached from the NL side.
     In the following, incorporating quantum fluctuations into $  f_{\psi} $( see Eqn.\ref{sep}
     ) and considering the repulsive coupling
     between $ \psi $ and $ n $ sector in Eqn.\ref{int}, we will determine the global phase diagram of  $^{4}He $.

{\bf 3. Two-component Quantum GL theory in the $ \psi $ sector: }
    In this section, we develop a two-component Quantum GL theory in the $ \psi $
    sector to replace Eqn.\ref{sfl} to describe the superfluid side of $^{4}He $.
    The superfluid is described by a complex order parameter $ \psi
    $ whose {\em condensation} leads to the Landau's quasi-particles. Although
    the bare $^{4}He $ atoms are strongly interacting, the Landau's quasi-particles are weakly interacting.
   The dispersion curve of superfluid state is shown in Fig.1 b which has both a phonon sector
   and a roton sector.  In order to focus on the low energy modes, we divide the spectrum into
   two regimes: the low momenta regime $ k < \Lambda $ where there
   are phonon excitations with linear dispersion and high momentum regime
   $ | k-k_{r} | < \Lambda \ll k_{r} $ where there is a roton minimum at the
   roton surface $ k=k_{r} $. We separate the complex order parameters $ \psi( \vec{x}, \tau)= \psi_{1}( \vec{x}, \tau) +
    \psi_{2}( \vec{x}, \tau) $
    into $ \psi_{1}( \vec{x}, \tau)= \int^{\Lambda}_{0}  \frac{d^{d} k}{ (2 \pi)^{d}
    } e^{i \vec{k} \cdot \vec{x} } \psi( \vec{k}, \tau) $ and $
    \psi_{2}( \vec{x}, \tau)= \int_{ | k-k_{r} | < \Lambda } \frac{d^{d} k}{ (2 \pi)^{d}
    } e^{i \vec{k} \cdot \vec{x} } \psi( \vec{k}, \tau) $ which stand for low
    energy modes near the origin and $ k_{r} $ respectively.
    For the notation simplicity, in the following, $ \int_{ \Lambda  } $ means $ \int_{ | k-k_{r} | < \Lambda  }
    $. The QGL action in the $ \psi $ sector in  the $ ( \vec{k}, \omega ) $ space becomes:
\begin{eqnarray}
   {\cal S}_{\psi} & =  & \frac{1}{2} \int^{\Lambda}_{0}  \frac{d^{d} k}{ (2 \pi)^{d} } \frac{1}{\beta} \sum_{i \omega_{n} }
      ( \kappa  \omega^{2}_{n} + t + K k^{2} )
      | \psi_{1}( \vec{k}, i \omega_{n} )|^{2}         \nonumber  \\
   & + & \frac{1}{2} \int_{ \Lambda  } \frac{d^{d} k}{ (2 \pi)^{d} } \frac{1}{\beta} \sum_{i \omega_{n} }
      (  \kappa \omega^{2}_{n} + \Delta_{r} + v_{r} ( k-k_{r} )^{2} )
      | \psi_{2}( \vec{k}, i \omega_{n} )|^{2}     \nonumber  \\
     & + &  u \int d^{d} x  d \tau |\psi_{1}( \vec{x}, \tau)+ \psi_{2}( \vec{x}, \tau)|^{4}+  \cdots
\label{sep}
\end{eqnarray}
    where  $ t \sim T-T_{SF} $ where $ T_{SF}
     \sim 2.17 K $ is the critical
    temperature of SF to NL transition at $ p= 0.05  \ bar $
    and $ \Delta_{r} \sim p_{c}-p $ where $ p_{c} \sim 25  \ bar $ is the critical
    pressure of SF to the SS transition at $ T=0 $.

{\bf 4. SF to SS transition and global phase diagram:}
     In the SF state, the Feymann relation between the Landau quasi-particle dispersion relation in the $ \psi $ sector
     and the static structure factor in the $ n $ sector holds:
\begin{equation}
    \omega( q ) = \frac{q^{2}}{ 2m S(q) }
\label{fey}
\end{equation}

      In the $ q \rightarrow 0 $ limit, $ S(q) \sim q $, $ \omega(q)
      \sim q $ recovers the $ \psi_1 $ sector near $ q=0 $.
      The first maximum peak in $ S(q) $ corresponds to the roton minimum in $
      \omega(q) $ in the $ \psi_2 $ sector, namely, $ k_{n}=k_{r} $.
      As one increases the pressure $ p $, the interaction
      $ u $ also gets bigger and bigger, the first maximum peak of $ S(q) $ increases,
      the roton minimum $ \Delta_{r} $ gets smaller and
      smaller. Across the critical pressure $ p=p_{c} $, there are
      two possibilities (1) The resulting
      solid is a commensurate solid, then $ < \psi >=0 $  (2) The resulting
      solid is an in-commensurate solid with vacancies even at $ T=0
      $ whose condensation leads to $ < \psi > \neq 0 $ \cite{and,ches,ander}.
      Which case will happen depends on the strength of the
      repulsive coupling $ v_{1} $ and will be analyzed in the next
      section. Case (1) is trivial, the SS phase in Fig.1 does not exist. In
      the following, we only focus on case (2).
      From Eqn.\ref{sl} and  Eqn.\ref{sep}, we can see
      that $ n $ and $ \psi_{2} $ have very similar propagators, so the lattice formation in $ n
      $ sector with  $ n(x)=  \sum_{\vec{G}} n_{\vec{G}} e^{i \vec{G} \cdot
       \vec{x} } $ and  the density wave formation in $ \psi_{2} $
      sector with  $ < \psi_{2}(\vec{x} ) >  = e^{i \theta_{2} } \sum^{P}_{m=1}
      \Delta_{m} e^{i  \vec{Q}_{m}  \cdot \vec{x} } $ where $ Q_{m}= k_{r} $
      happen simultaneously. The $ \psi_{2} $ sector alone is
      described by  $ n=2 $ component $ (d+1,d) $ with $ d=3 $
      Lifshitz action \cite{tom}.
      The repulsive coupling in Eqn.\ref{int} $ v_{1} > 0 $ simply
      shifts the DW by suitable constants along the three unit vectors in
      the direct lattice. These constants will be determined in the
      next section for different  $ n $ lattices.
      Namely, the SS state consists of two inter-penetrating lattices formed by
      the $ n $ lattice and the $\psi $ superfluid density wave (SDW).

      Combining the roton condensation picture in this section with the
      results in section 2,
      we can sketch the following global phase diagram of  $^{4}He $ in case (2).

\vspace{0.25cm}

\epsfig{file=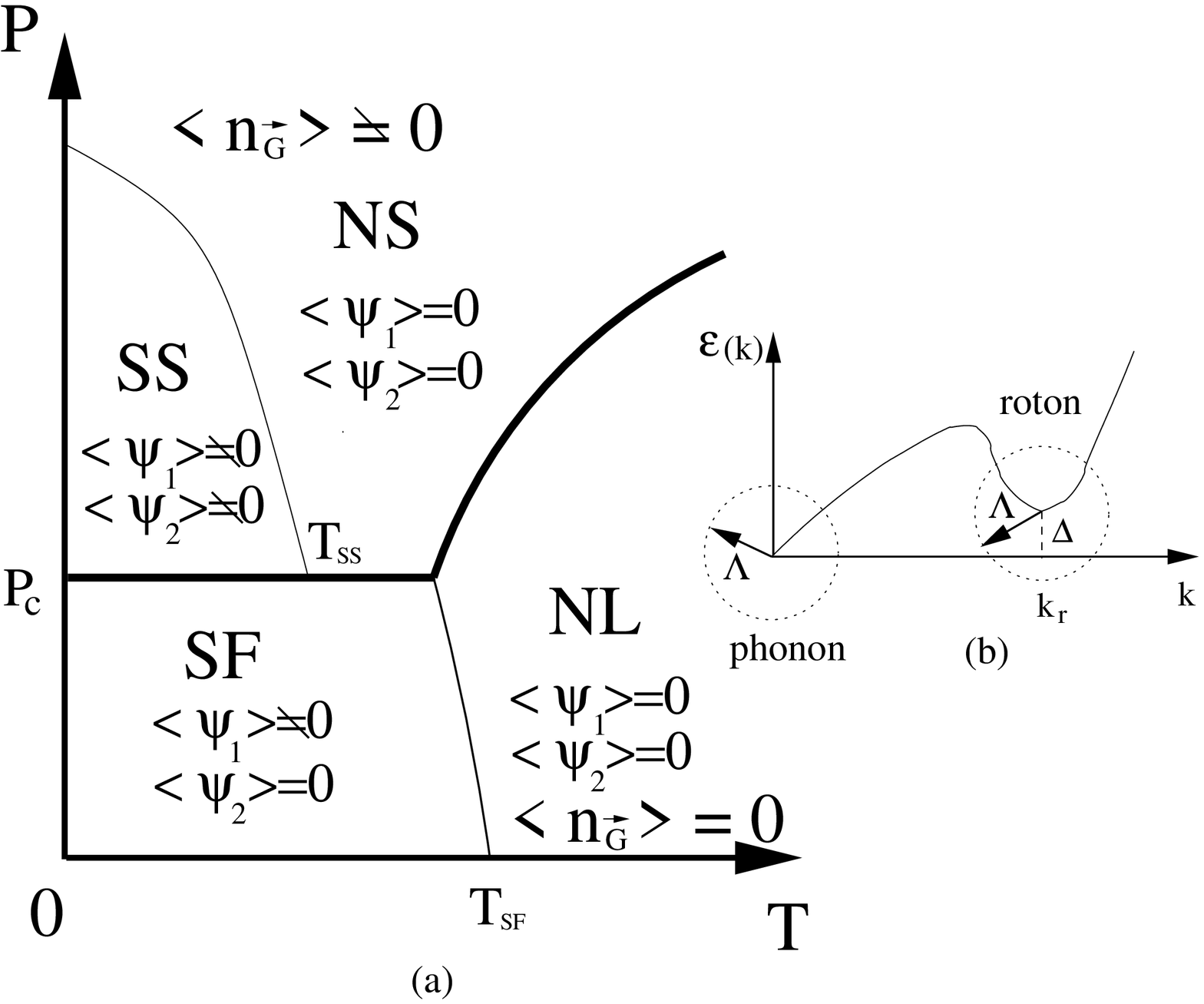,width=3.0in,height=2.0in,angle=0}

\vspace{0.25cm}

{\footnotesize {\bf Fig.1:} (a) The phase diagram in case (2).
   $ T $ controls thermal fluctuations, while $ P $ tunes quantum fluctuations.
   The SS exists only when the repulsive coupling  $ v_{1} $ in Eqn.\ref{int} is sufficiently small.
   Thick ( thin ) lines are 1st ( 2nd ) order transitions. The critical
   temperatures of NL to SF and NS to SS transitions
   drop slightly as the pressure $ p $ increases because of the quantum fluctuations \cite{un}.
   (b) The separation  of low ( phonon ) and high ( roton ) momenta
   regime in the SF.}

\vspace{0.25cm}

   {\bf 5. The NS to SS transition: }
   In this section, we approach the SS phase from the NS
   side and determine its lattice structure.
   In the NL, the BEC happens in the $ \psi_{1} $ sector at $ k=0 $, $ \psi_{2}
   $ has a large gap and can be simply integrated out.
   In the NS, $ \psi $ stands for the vacancies due to large zero point quantum fluctuations \cite{and,ches,ander}.
   Due to the $ n $ lattice formation, the mass of $ \psi_{1} $ was
   increased to $ t^{NS}_{\psi_{1}} = t +  v^{NS}_{1} n_{0} = T-T_{SS} $.
   Subtracting it from $ t^{NL}_{\psi_{1}} = t +  v^{NL}_{1} n_{0}=T-T_{SF}
   $ leads to $ T_{SF}-T_{SS}=n_{0} \Delta v_{1} $.
   For simplicity, one can set $ v^{NL}_{1} =0 $.
   If $ v_{1} $ is larger than a critical value $ v > v_{c1} $, then
   $ T_{SS} $ is suppressed to zero. This is the case (1) discussed in the last section.
   The SS disappears from the phase diagram Fig.1. The resulting
   solid is a commensurate solid.
   In the following, we only focus on $ v_{1} < v_{c1} $,  so
   the SS state exists in the Fig.1 with $ T_{SS} >0 $ which is the case (2) in the last section.
   The resulting solid is an in-commensurate solid with
   vacancies even at $ T=0 $ whose condensation leads to the formation of the SS. Then
   the temperature shift $  \Delta T = T_{SF}- T_{SS} $ is an effective
   experimental
   measure of the repulsive coupling $ v_{1} $ in Eqn.\ref{int}.
   In  the presence of the periodic potential of $ n(x) $ lattice,
   $ \psi $ will form a Bloch wave, the $ u $ self-interaction in the $ \psi $ sector in
   Eqn.\ref{sep} will favor extended Bloch wave over
   strongly localized Wannier state. In principle, a full energy
   band calculation incorporating the interaction $ u $ is
   necessary to get the energy bands of $ \psi $. Fortunately,
   qualitatively physical picture can be achieved without such a detailed energy band calculation.
   In the following, substituting the ansatz $ < \psi_{1}(\vec{x} )
   >= a e^{i\theta_{1} } $ and $  < \psi_{2}(\vec{x} ) >  = e^{i \theta_{2} } \sum^{P}_{m=1}
   \Delta_{m} e^{i  \vec{Q}_{m}  \cdot \vec{x} } $ where $ Q_{m}= Q $ into Eqn.\ref{int},
   we study the effects of  $ n $ lattice  on $ \psi= \psi_{1}+ \psi_{2} $.
   In order to get the lowest energy ground state, we must consider
   the following 4 conditions: (1)  because any complex $ \psi $ ( up to a global phase ) will lead to local
   supercurrents which is costy, we can also take $ \psi $ to be
   real, so $ \vec{Q}_{m} $ have to be paired as anti-nodal points. $ P $ has to be even (2)
   as shown from the Feymann relation Eqn.\ref{fey}, $ \vec{Q}_{m}, m=1,\cdots,P $ are
   simply $ P $ shortest reciprocal lattice vectors, then translational symmetry of the lattice dictates that
   $ \epsilon( \vec{K}=0 ) = \epsilon( \vec{K}=\vec{Q}_{m}) $,
   $ \psi_{1} $ and $ \psi_{2} $ have to condense at the same time. (3) The point group symmetry of the lattice
   dictates $ \Delta_{m} =\Delta $ and is real (4)  The
   repulsive interaction $ v_{1} > 0
   $ favors $ \psi(x=0) =0 $, so the Superfluid Density Wave (SDW) $ \rho=|\psi|^{2} $ can avoid the $ n $ lattice as much as
   possible. It turns out that the 4 conditions can fix the relative phase and magnitude of $ \psi_{1} $ and
   $\psi_{2} $ to be $ \theta_{2}=\theta_{1}+ \pi, \Delta= a/P $, namely,
   $ \psi= a e^{i \theta} ( 1- \frac{2}{P} \sum^{P/2}_{m=1} \cos \vec{Q}_{m} \cdot \vec{x} ) $.
   In this state, the crystal momentum $ \vec{k}=0 $ and  the Fourier components are $ \psi( \vec{K}=0)=
   a, \psi( \vec{K}= \vec{Q}_{m} )= - a/P $ which oscillate in sign and decay in magnitude. In principle, higher
   Fourier components may also exist, but they decay very rapidly, so
   can be neglected without affecting the physics qualitatively.
   In the following, we will discuss 4 common lattices
   $ sc,  fcc, bcc , hcp $ respectively.

  (a) $ P=6 $: $ \vec{Q}_{i}, i=1,2,3,4,5,6 $ are the 6 shortest
   reciprocal lattice vectors generating a cubic lattice. The maxima of the DW $ \psi_{max}=2 a $ appear exactly in the middle of lattice
   points at the 8 points $ \vec{a}=  \frac{1}{2} ( \pm \vec{a}_{1} \pm
   \vec{a}_{2} \pm \vec{a}_{3} ) $. They form the dual lattice of the cubic lattice which is
   also a cubic lattice.

  (b)  $ P=8 $:  $ \vec{Q}_{i}, i=1,\cdots,8 $ form  the 8 shortest
   reciprocal lattice vectors generating a $ bcc $ reciprocal
   lattice which corresponds to a $ fcc $ direct lattice. The field is
   $ \psi ( \vec{x})=a[1- \frac{1}{4}( \cos \vec{Q}_{1} \cdot \vec{x} + \cos \vec{Q}_{2} \cdot \vec{x}
   +  \cos \vec{Q}_{3} \cdot \vec{x} +  \cos \vec{Q}_{4} \cdot \vec{x}  )] $,
   The local superfluid density $ \rho^{l}_{DW} = |  \psi(\vec{x} )
   |^{2} $. The maxima of the DW $ \psi_{max}=2 a $ appear in all the edge centers such as $ (1/2,0,0) $ etc. and
   the centers of any cube such as $ (1/2,1/2,1/2) $.

  (c)  $ P=12 $:  $ \vec{Q}_{i}, i=1,\cdots,12 $ form  the 12 shortest
   reciprocal lattice vectors generating a $ fcc $ reciprocal
   lattice which corresponds to a $ bcc $ direct lattice. The field is
    $ \psi( \vec{x}) = a[1- \frac{1}{6}( \cos \vec{Q}_{1} \cdot \vec{x} + \cos \vec{Q}_{2} \cdot \vec{x}
     +  \cos \vec{Q}_{3} \cdot \vec{x} +  \cos \vec{Q}_{4} \cdot \vec{x}+
      \cos \vec{Q}_{5} \cdot \vec{x} +  \cos \vec{Q}_{6} \cdot \vec{x}   )]
      $.  The maxima of the DW  $ \psi_{max}= 4/3 a $ appear along
   any square surrounding the center of the cube such as $ (1/2,\beta,0)
   $ or $ (1/2,0,\gamma) $ etc. Note that $^{4}He $ in Vycor glass takes a $ bcc $ lattice

   (d) Unfortunately, a spherical
   $ k_{r} = Q $ surface can not lead to lattices with {\em different} lengths
   of primitive reciprocal lattice vectors such as a $ hcp $ lattice. This is
   similar to the classical liquid-solid transition described by Eqn.\ref{sl} where a single
   maximum peak in the static structure factor can not lead to a $ hcp $ lattice \cite{tom}.
   Here we can simply take the experimental fact that $ n $ forms a $ hcp
   $ lattice without knowing how to produce such a lattice from a GL
   theory Eqn.\ref{sl}. Because for an idea $ hcp $ lattice $ c/a =\sqrt{8/3}
   $, an $ hcp $ lattice has 12 nearest neighbours, so its local
   environment may resemble that of an $ fcc $ lattice.
   We expect the physics ( except the weak anisotropy of NCRI in the $ hcp $ lattice discussed
   in \cite{un} ) is qualitatively the same as that in $ fcc
   $ direct lattice.

    Let's look at the prediction of our theory on X-ray
    scattering from the SS. For simplicity, we take the $ sc $
    lattice to explain the points. The other lattices will be discussed in \cite{un}. For a lattice with $ j=1, \cdots, n $ basis located
    at $ \vec{d}_{j} $, the geometrical structure factor at the reciprocal lattice vector $  \vec{K} $
    is $ S ( \vec{K} )= \sum^{n}_{j=1} f_{j}( \vec{K} ) e^{i \vec{K} \cdot \vec{d}_{j}}  $ where $ f_{j}
    $ is the atomic structure factor of the basis at $ \vec{d}_{j}
    $.  The X-ray scattering amplitude $ I( \vec{K}) \sim | S ( \vec{K} ) |^{2} $. For the SS in the $ sc $ lattice,
    $ \vec{K} = \frac{2 \pi}{ a } ( n_{1} \vec{i} + n_{2} \vec{j} + n_{3} ) \vec{k},
    \vec{d}_{1}=0, \vec{d}_{2}= \frac{a}{2} ( \vec{i} + \vec{j} + \vec{k} ) $, then $ S( \vec{K})=1+ f (-1)^{ n_{1} + n_{2} + n_{3} } $
    where $ f \sim \rho^{max}_{s} \sim  a^{2} $. It is $ 1+ f $ for even $
    \vec{K} $ and $ 1 - f $ for odd $ \vec{K} $. In fact, due to large zero-point motion, any X-ray
    scattering amplitude $ I (\vec{K}) $ will be diminished by a Debye-Waller factor.
     As shown in \cite{un}, the lattice phonon modes $ \vec{u} $ in
    $ \psi $ are {\em locked} to those of $ n $, so there is a {\em common}
    Debye-Waller factor $ \sim e^{- \frac{1}{3} K^{2} <u^{2}> } $
     for both even and odd $ \vec{K} $. We conclude that the {\em elastic } X-ray
    scattering intensity from the SS has an additional modulation
    over that of the NS. The modulation amplitude is proportional
    to the maxima of the superfluid density $ \rho^{max}_{s} \sim  a^{2} $ which is
    the same as the NCRI observed in the PSU's torsional oscillator experiments.
    Unfortunately, so far, the X-ray scattering data is limited to
    high temperature $ T> 0.8 K > T_{SS} $ \cite{simmons}. X-ray scattering experiments on lower
    temperature $ T< T_{SS} $ are needed to test this prediction.

   The results achieved in this
   section indeed confirm Fig.1 achieved from the roton condensation picture in the last
   section. A low energy effective action involving the superfluid
   phonon $ \theta $, the lattice phonons $ \vec{u} $ and their
   couplings will appear in a future publication.

{\bf 6. The vortices in the SS: }
      In the sectors of $ \psi $, there are topological defects in the phase winding  of $ \theta
      $ which are  vortices.  At $ T \ll T_{SS} $, the vortices  can only appear in tightly bound
      pairs. However, as $ T \rightarrow T^{-}_{SS} $, the vortices start to
      become liberated, this process renders the total NCRI to vanish above $ T > T_{SS} $
      in the NS state. In the SF phase, a single vortex energy
      costs a lot of energy $ E^{SF}_{v} = \frac{ \rho^{SF}_{s} h^{2} }{
      4 \pi m^{2} } ln \frac{R}{\xi_{SF}} $ where $ m $ is the mass of $
      ^{4}He $ atom, $ R $ is the system size and $ \xi_{SF} \sim a  $ is the core size of
      the vortex. This energy determines the critical velocity in SF
      $ v^{SF}_{c} > 30cm/s $. In the SS state, because in the center of the SS vortex core,
      $ \psi= 0 $, so the vortices prefer to sit on a lattice site of the $ n $ lattice.
      Because the long distance behavior of SS is more or less the same as
      SF, we can estimate its energy $ E^{SS}_{v} = \frac{ \rho^{SS}_{s} h^{2} }{
      4 \pi m^{2} } ln \frac{R}{ \xi_{SS} } $. We expect the core size of a supersolid vortex
      $ \xi_{SS} \sim 1/\Lambda  \gg 1/k_{r} \sim a  \sim  \xi_{SF} $. so inside the  SS vortex core,
      we should also see
      the lattice structure of $ n $. In fact, we expect that $ \xi_{SS} $ is of the order of
      the average spacing between the vacancies in the SS. It is interesting to see if neutron or light scattering  experiments can test this prediction.
      Compared to $ E^{SF}_{v} $,
      there are two reductions, one is the superfluid density,
      another is the increase of the vortex core size $ \xi_{SS} \gg \xi_{SF} $. These two factors contribute
      to the very low critical velocity $ v^{SS}_{c} \sim 30 \mu m/s $.

{\bf 7. Discussions on PSU's experiments: }
    Although the NS to SS transition is in the same universality
    class as the NL to SF one \cite{dorsey}, it may have quite different
    off-critical behaviours due to the SDW structure in the SS state.
    We can estimates the critical regime of the NS to SS transition from the Ginsburg
    Criterion $ t^{-1}_{G} \sim \xi^{3}_{SS} \Delta C $ where $ \Delta
    C $ is the specific jump in the mean field theory. Because of the
    cubic dependence  on $ \xi_{SS} $, large $ \xi_{SS} $ leads
    to extremely narrow critical regime, the 3D $ XY $ critical behavior is essentially
    irrelevant, instead mean field Gaussian theory should apply.
    This fact can be used to address two experimentally observable effects of the He3 impurities.
    The first one was already observed by the PSU experiments and the second one is
    underway \cite{private}. (1) The unbinding transition temperature $ T_{SS} $ is determined by
    the pinning due to impurities instead of by the logarithmic interactions
    between the vortices.
    So the $^{3}He $ impurities effectively pin the vortices and raise the unbinding critical
    temperature $ T_{SS} $.
    On the other hand, $^{3}He $ impurities  will  certainly decrease the
    superfluid density in  both the $ \psi_{1} $ and $ \psi_{2} $ sector just like$^{3}He $ impurities
    decreases superfluid density in the $^{4}He $ superfluid. (2)
    The mean field specific heat jump at $ T=T_{SS} $ may be smeared by the presence of $^{3}He $
    impurities.

{\bf 8. Conclusions:} The GL theory developed in this paper leads
    to a global and unified picture of $^{4}He $ physics
    at any temperature and pressure. If the repulsive coupling
    in Eqn.\ref{int} is sufficiently small, the SS state
    becomes a ground state at zero temperature. The solid at
    high pressure is an incommensurate solid with zero point
    fluctuations generated vacancies whose condensation leads to the
    formation of the SS. Assuming this is indeed the case,
    we investigate the SS state from both the SF and the NS side
    and find completely consistent description of the properties of
    the SS state. We found that the SF to the SS transition is a first order
    transition driven by the collapsing
    of roton minimum in the SF side, while the NS to SS
    transition is described by a 3d $ XY $ model with much {\em narrower} critical regime than the NL to SF
    transition.
   The SS state is a uniform two-component phase consisting
   of a normal lattice plus a
    commensurate superfluid density wave (SDW). The SDW in the SS
    state leads to a modulation on the X-ray
    scattering intensity over that of the NS. The modulation amplitude is proportional
    to the NCRI observed in the PSU's torsional oscillator experiments.

  {\bf Acknowledgement}

    I am deeply indebted to Tom Lubensky for many insightful and critical comments on the
    manuscript. I thank  P. W. Anderson, T. Clark, M. Cole, B.
    Halperin, J. K. Jain, T. Leggett, G. D. Mahan, especially Moses Chan for helpful
    discussions.  The research at KITP was supported by the NSF grant No.
    PHY99-07949.

\end{multicols}
\end{document}